\documentclass[10pt]{article}

\usepackage{amsfonts, amssymb}
\usepackage{amsthm}
\usepackage{amsmath}

\newcommand{\jdot}{\bullet}

\newcommand{\beq}{\begin{equation}}
\newcommand{\eeq}{\end{equation}}
\newcommand{\beqa}{\begin{eqnarray}}
\newcommand{\eeqa}{\end{eqnarray}}

\newcommand{\p}{{\mathfrak p}} 
\newcommand{\g}{{\mathfrak g}}

\renewcommand{\k}{{\mathfrak k}}

\newcommand{\ran}{\mbox{ran}}

\newcommand{\1}{\mbox{{\bf 1}}}
\newcommand{\tr}{\mbox{tr}}

\newcommand{\Aut}{\mbox{Aut}}

\newcommand{\A}{{\mathfrak A}}

\newcommand{\C}{{\mathcal C}} 
  
\newcommand{\cc}{{\cal C}}  
\newcommand{\D}{{\cal D}} 

\newcommand{\E}{{\bf E}}  
\newcommand{\F}{{\bf F}} 

\newcommand{\G}{{\cal G}} 

\renewcommand{\H}{{\mathbf H}}

\newcommand{\cl}{{\cal L}}    
\renewcommand{\O}{{\cal O}}   
\newcommand{\R}{\mathbb{R}}
\newcommand{\V}{{\bf V}}      

\newtheorem{theorem}{Theorem}[]

\newtheorem{lemma}[theorem]{Lemma}
\newtheorem{proposition}[theorem]{Proposition}
\newtheorem{corollary}[theorem]{Corollary}

\newtheorem{definition}[theorem]{Definition}

\theoremstyle{remark}

\title{Local tomography and the Jordan structure of quantum theory} 
  \author{Howard
  Barnum\footnote{Department of Physics and Astronomy, University of
    New Mexico; {\tt hnbarnum@aol.com, hbarnum@unm.edu}}~\footnote{Stellenbosch Institute for Advanced Study (STIAS), 
Wallenberg Research Centre at Stellenbosch University, 
Marais Street, Stellenbosch 7600, South Africa    
    } ~and Alexander
  Wilce\footnote{Department of Mathematics, Susquehanna University;
    {\tt wilce@susqu.edu}}} 

\begin{document}

\maketitle

\begin{abstract} 
Using a result of H. Hanche-Olsen, we show that (subject to fairly
natural constraints on what constitutes a system, and on what
constitutes a composite system), orthodox finite-dimensional complex
quantum mechanics with superselection rules is the only non-signaling
probabilistic theory in which (i) individual systems are Jordan
algebras (equivalently, their cones of unnormalized states 
are homogeneous and self-dual), (ii) composites are locally
tomographic (meaning that states are determined by the joint
probabilities they assign to measurement outcomes on the component
systems) and (iii) at least one system has the structure of a
qubit. Using this result, we also characterize finite dimensional
quantum theory among probabilistic theories having the structure of a
dagger-monoidal category.
\end{abstract}

\section{Introduction and background} 

One of the oldest foundational problems besetting quantum mechanics is
to provide a clear motivation for its probabilistic apparatus --- in 
particular, for the representation of observables of a quantum system by the self-adjoint elements of 
a $C^{\ast}$ algebra. {\em Why} should outcomes of measurements give rise to anything so nicely
structured as a $C^{\ast}$-algebra --- or any algebra at all, for that
matter? In particular, what operational meaning can we give to the
product of two non-commuting observables, when these cannot
simultaneously be measured, and when, indeed, this product is not  
self-adjoint?  

\subsection{Jordan algebras}
In an early attempt to address this question, Pascual Jordan 
\cite{Jordan}  proposed in 1932 that the observables
associated with a finite-dimensional physical system 
should constitute what is now called a formally real
Jordan algebra.  A Jordan algebra is a finite-dimensional real vector space $\E$
equipped with a commutative bilinear operation $\jdot : \E \times \E
\rightarrow \E$ satisfying the {\em Jordan identity}
\[a^2 \jdot (a \jdot b) = a \jdot (a^2 \jdot b)\]
for all $a, b \in \E$ (where $a^2 := a \jdot a$).  Jordan algebras are
naturally equipped with a bilinear trace form, which induces a
symmetric, nondegenerate bilinear form $(a,b) \mapsto \langle a, b
\rangle := \tr(a \bullet b)$.
If this is an inner product (that is, positive-definite), one calls $\E$ {\em Euclidean}. In 
finite dimensions, this is equivalent to Jordan's condition of {\em formal reality}: that $a^2 + b^2 = 0 \ \Rightarrow x = y = 0$.
Two years later, Jordan, von Neumann, and Wigner \cite{JvNW}
classified such algebras as being either (i) self-adjoint parts of
(real, complex or quaternionic) matrix algebras, under the
anti-commutator $x \jdot y = (xy + yx)/2$, (ii) so-called spin
factors, or (iii) the self-adjoint part of the 3-by-3 matrix algebra
over the octonions, or direct sums of these. Thus, the assumption that
the space of observables of a physical system is a Euclidean Jordan
algebra does bring one very close to finite-dimensional quantum
mechanics.


\subsection{General probabilistic theories} 

An early objection to the Jordan-algebraic approach is that it does
not generalize easily to the infinite-dimensional setting required for
full-blown quantum mechanics. However, in recent years, with the
growing importance of quantum information theory, finite-dimensional
quantum theory is coming to be viewed as an important subject in its
own right, and is even regarded (in some quarters) as being possibly
more fundamental than more traditional, infinite-dimensional QM.
Moreover, the introduction of the notion of JB-algebra has turned out
to provide a fairly satisfactory generalization to infinite dimension.

Leaving this issue to one side, another, and more basic, objection is 
that the Jordan product has no
clearer an operational interpretation than the $C^{\ast}$-algebraic 
product. Later work has tended to start with a much more general (but conceptually 
much more transparent) framework \cite{Mackey, Holevo, BBLW07}, in which a 
physical --- or, more broadly, probabilistic --- system is represented by an {\em order-unit space}. This is an 
ordered real vector space $\E$, with positive 
cone $\E_+$, equipped with a distinguished element $u \in \E_+$, called the {\em order unit}, such that for every $a \in \E_+$, $ta \leq u$ for some $t > 0$. Possible measurement 
outcomes associated with the system are identified with {\em effects}, that is, vectors $a \in \E_+$ with $a \leq u$. States of the system 
are identified with positive linear functionals $\alpha : \E \rightarrow \R$ with $\rho(u) = 1$. If $a$ is an effect, then $\alpha(a)$ is understood to be the probability 
that $a$ will occur (if measured) when the state $\alpha$ obtains. Physical processes acting on a system, or between two systems, can then be represented very naturally 
by positive linear mappings between the associated ordered linear spaces, and one can define a {\em probabilistic theory} to be a category of such spaces and mappings. 

Within this very general setting (which we review in greater detail in
Section 2), one can hope to find illuminating characterizations of
quantum theory, that is, theorems that single out QM --- particularly,
complex QM --- as the unique probabilistic theory satisfying one or
more reasonable constraints. This was the goal, explicit or tacit, of
a great deal of foundational work in quantum theory from roughly the
1950s to the late 1970s \cite{Mackey, Ludwig}.  With the emergence of
quantum information theory, this project has enjoyed a strong revival,
with a distinctive focus on finite-dimensional systems, and an
emphasis on composite systems \cite{Hardy, Rau, Goyal, CDP,
Dakic-Brukner, Masanes}, The cited papers all come close to, or indeed
succeed in, deriving finite dimensional QM from simple
axioms. However, many \cite{Hardy, Rau, Dakic-Brukner, Masanes} make
use of a strong uniformity principle, namely, that all systems having
the same information-carrying capacity (as variously defined) are
isomorphic; others place strong constraints on the representation of
sub-systems \cite{Hardy, CDP}.  We hope to avoid both kinds of
assumptions.

\subsection{Homogeneity and self-duality}
A different approach, which we have pursued in \cite{BGW, BDW10,
Wilce09, Wilce11}, is to exploit the classical correspondence between
Jordan algebras and homogeneous self-dual cones. This is reviewed in
more detail below, but, briefly: The positive cone $\E_+$ of an
ordered vector space $\E$ is {\em homogeneous} iff the group of
order-automorphisms\footnote{That is, positive linear bijections  
having positive inverses} of $\E$ acts transitively on the {\em interior} of
$\E_+$, and {\em self-dual} iff there exists an inner product on $\E$
such that
\[\E_+ = \E^{+} := \{ a \in \E | \langle a, b \rangle \geq 0 \ \forall b \in \E_+\}.\]

{\bf Theorem (Koecher \cite{Koecher}, Vinberg \cite{Vinberg}):} {\em
Let $\E$ be a finite-dimensional order-unit space with a
homogeneous, self-dual (HSD) cone $\E_+$. Then there exists a unique
bilinear operation $\bullet : \E \times \E \rightarrow \E$ making $\E$
into a Euclidean Jordan algebra with unit $u$ and cone of squares equal to 
$\E_+$.}

In \cite{BGW}, we observed that a simple purification or dilation
principle is enough to guarantee that the cone of states of a physical
system is homogeneous and {\em weakly} self-dual, i.e., $\E^{\ast}_{+}
\simeq \E_{+}$. However, the distinction between weak self-duality and
self-duality is significant, so this result still leaves us with two
questions: first, why the state cone ought to be self-dual, and,
secondly, how to rule out, or to make room for, the various
alternatives to complex QM allowed by the Jordan-von-Neumann-Wigner
classification.

In this paper, we bracket the first question (to which several
possible answers have been suggested; see \cite{BGW, BDW10,
Muller-Ududec, Wilce09, Wilce11}) and concentrate on the second. We
consider a probabilistic theory in which (i) individual systems are
represented by homogeneous, self-dual models --- equivalently, by
formally real Jordan algebras --- and ask when these must in fact be
standard {\em quantum} models, i.e, the self-adjoint parts of complex
matrix algebras.

\subsection{Composites of homogeneous, self-dual systems} \label{sec: composites of HSD}
As it happens, a {\em nearly} off-the-shelf answer is available. It
has been known at least since \cite{Araki} that complex QM is
distinguished from its real analogue by a property called {\em local
tomography}, which requires that the joint state of a composite system
be completely determined by the joint probabilities assigned to
observables on the two component systems.\footnote{It is also observed
in \cite{Araki} that in the quaternionic analogue of complex quantum
theory, the most obvious candidate for the state space of a composite
of $m$-dimensional and $n$-dimensional quaternionic systems, namely
the $mn \times mn$ dimensional positive semidefinite (PSD)
quaternionic matrices, suffers from difficulties in even identifying
the product effects necessary to a locally tomographic
composite---indeed, its dimension is smaller than the product of the
dimensions of the spaces spanned by the 
$m \times m$ and by the $n \times n$ quaternionic positive semi-definite 
matrices.}  In \cite{Hanche-Olsen}, H. Hanche Olsen made a similar  
point regarding Jordan algebras:  

{\bf Theorem (Hanche-Olsen, \cite{Hanche-Olsen}):} {\em Let $\E_2$ be
  the Jordan algebra of hermitian $2 \times 2$ complex matrices, i.e.,
  the Jordan algebra corresponding to a single qubit.  Let $\E$ be any
  JB algebra (in finite-dimensions, the same thing as a Euclidean
  Jordan algebra), and suppose that the vector space $\E \otimes \E_2$
  carries a Jordan product satisfying \begin{equation} \label{eq:
      Hanche-Olsen condition}(a \otimes \1) \bullet (b \otimes v) = (a
    \bullet b) \otimes v \ \text{and} \ (\1 \otimes v) \bullet (a
    \otimes w) = a \otimes (v \bullet w). \end{equation} 
    for all $a,b \in \E$ and all $v,w \in \E_2$. Then $\E$ is
  the Hermitian part of a $C^{\ast}$-algebra.}

Of course, absent a direct physical or operational interpretation of
the Jordan product, Hanche-Olsen's condition (\theequation) calls for
some further motivation.  In Section 4, we show that in the context of
composites of probabilistic models, (\theequation) follows from {\em
  local tomography} --- the condition that the joint state of a
composite system is determined by the joint probabilities it assigns
to outcomes of measurements on the two component systems --- plus the
condition that the self-dualizing inner product on a composite system
can be chosen so as to factor into a product of self-dualizing inner
products for the component systems. We call a theory satisfying the
latter condition {\em factorizably self-dual}. By a {\em factorizably
  HSD theory}, we mean a probabilistic theory in which every system is
homogeneous and factorizably self-dual.  Hanche-Olsen's result then
yields

\begin{proposition} \label{prop: main} Let $\cc$ be any factorizably HSD 
  probabilistic theory in which (i) every pair of systems $A$ and $B$
  admit a locally-tomographic composite system, $AB$, still belonging
  to $\cc$, and (ii) there exists a qubit.  Then all systems in $\cc$
  are self-adjoint parts of complex matrix algebras.
\end{proposition}

The factorizability assumption can itself be further motivated. In
particular, it is automatically satisfied given two very weak and
natural conditions, namely, that each component system support a {\em
  uniform} (or {\em maximally mixed}) state, and that every basic
measurement outcome have probability one in {\em some} state. Given
these assumptions, finite-dimensional QM is completely characterized
among finite-dimensional HSD theories by conditions (i) and (ii)
above.

Proposition 1 has an important consequence for the categorical
formulation of quantum theory in terms of dagger-monoidal categories
\cite{Abramsky-Coecke, Baez, Selinger}. Let $\C$ be a dagger-monoidal
category whose objects are order-unit spaces, with the set of
morphisms between any two objects being a cone of positive linear
mappings between these spaces, with tensor unit $I \simeq \R$.  If
$\C(I,A) \simeq A$, then each object $A$ is equipped with a canonical
bilinear form, namely $\langle a, b \rangle = a^{\dagger} \circ b$,
which factors on tensor products. If this is an inner product, and the
group of invertible elements of $\C(A,A)$ acts homogeneously on $A$,
then the positive cone of $A$ is self-dual. Thus, if tensor products
in $\C$ are locally tomographic, non-signaling composites, and if $\C$
contains a qubit, then every order-unit space $A \in \C$ is the hermitian part of
a $C^{\ast}$ algebra. 

The balance of this note supplies the proof of Proposition 1, along
with enough technical background to make the exposition
self-contained. In Section 2, we give a more detailed sketch of the
general probabilistic framework described above, and discuss the
structure of models associated with formally real Jordan algebras.  In
Section 3, after discussing composite systems in general, we study
locally tomographic composites of Jordan-algebraic sytems, and prove
Proposition 1. In Section 4, we reconsider these ideas in the context
of a dagger-monoidal category of probabilistic models. Section 5
offers a few concluding remarks, questions, and speculative
suggestions.

\section{Probabilistic Models and Theories} 

In this section we provide a quick review of the framework for
generalized probability theory that we shall use. This is fairly
standard, with a history going back ultimately to the work of Mackey
in the 1950s. The precise machinery we use combines ideas borrowed
from \cite{Davies-Lewis, Holevo, Foulis-Randall}, here specialized to
finite-dimensional systems.

\subsection{States, Effects and Processes}

In its very simplest formulation, classical probability theory
concerns an ``experiment" --- a single, discrete set $E$ of mutually
exclusive possible outcomes, and probability weights thereon.  A
particularly simple (and, conceptually, very conservative)
generalization of classical probability theory begins with the idea
that one may be faced with a choice of experiments.

\begin{definition} {\em A {\em test space} is a family $\A$ of non-empty sets, called {\em tests}, construed as the outcome-sets associated with various experiments, measurements, or other operations. The {\em outcome space} of $\A$ is the set $X := \bigcup \A$ of all outcomes arising from any test $E \in A$. A {\em state}, or {\em probability weight}, on $\A$ is a mapping $\alpha : X \rightarrow [0,1]$ summing to unity on each $E \in \A$ --- in other words, $\alpha$ is a simultaneous (and non-contextual) assignment of a probability weight to each test. }\end{definition}

{\bf Examples:} (i) A discrete classical test space is one of the form
$\{E\}$, that is, one that contains only a single test. (ii) One can also
consider the test space consisting of finite (respectively, countable)
partitions of a measurable space $S$ by measurable subsets; in this
case, the states correspond exactly to finitely additive
(respectively, countably additive) probability measures on $S$.  (iii) The standard
test space in quantum theory is the collection of maximal sets of
pairwise orthogonal, rank-one projection operators on a Hilbert space
$\H$. Gleason's Theorem tells us (for $\dim(\H) > 2$) that all
probability weights on this test space are implemented by density
operators, according to the ``Born rule".

{\bf States and Effects} Given a test space $\A$, it is often reasonable to consider a
restricted state space $\Omega$. (For instance, given a qubit, we
typically restrict attention to those states given by density
operators, rather than allowing the various discontinuous states that
would otherwise be allowed by the very loose combinatorial structure
of $\F_2$.) Plausibly, $\Omega$ should be both convex and closed with
respect to outcome-wise convergence --- hence, compact as a subset of
$[0,1]^{X}$. It should also be rich enough to separate outcomes, in
the sense that if $x, y \in X$ and $\alpha(x) = \alpha(y)$ for all
$\alpha \in \Omega$, then $x = y$. 
We can now associate to every $x \in X$ the corresponding evaluation functional $\alpha \mapsto
\alpha(x)$ in $\R^{\Omega}$. Let $\E$ denote the span of $X$ in
$\R^{\Omega}$. We shall say that the pair $(\A,\Omega)$ is
{\em finite-dimensional} iff $\E$ is finite dimensional.

Now define a cone in $\E$ by setting $\E_+ = \{ \sum_{i} t_i x_i | t_i
\geq 0, x_i \in X\}$. Let $u$ denote the unit functional $u(\alpha)
\equiv 1$; then $\sum_{x \in E} x = u$ for every test $E \in \A$. In
particular, $x \leq u$ for every $x \in X$. It follows that $u$ is an
order-unit for $\E$. If $\alpha \in \E^{\ast}$ is any normalized
positive functional, i.e, $\alpha(a) \geq 0$ for $a \in \E_+$ and
$\alpha(u) = 1$, then we obtain a state on $\A$ by restriction to
$X$. The set of states arising in this way defines a compact convex
set $\widehat{\Omega} \supseteq \Omega$. Call $\Omega$ {\em state-complete} iff
$\widehat{\Omega} = \Omega$. It is reasonable to assume, and we shall assume
here, that {\bf \emph{all state spaces are state-complete}}.  So
for the remainder of the paper, ``state'' means ``element of
$\widehat{\Omega}$''.

{\bf Processes} Any test space $\A$ is associated with a group of {\em
  symmetries}, i.e., bijections $g : X \rightarrow X$ with $gE \in \A
\ \leftrightarrow E \in \A$ for all $E \subseteq X$. This group is
compact in $\R^{X}$, and acts on $\E$ by positive, unit-preserving
linear automorphisms. Just as it may be reasonable to restrict the set
of states, it may be desirable to consider a restricted set of
symmetries. More generally, we may wish to identify a semigroup of
``physical processes". Such processes should surely map normalized states to
possibly sub-normalized states, preserving convex combinations. Thus,
we might represent a physical process by a positive mapping $\phi :
\E^{\ast} \rightarrow \E^{\ast}$, with $u(\phi(\alpha)) \leq
u(\alpha)$ for all $\alpha \in \E^{\ast}_{+}$.  We interpret
$u(\phi(\alpha))$ as the {\em probability} that $\phi$ occurs when the
initial state is $\alpha$.

If $\phi : \E^{\ast} \rightarrow \E^{\ast}$ is a physical process, there will be a dual process $\tau = \phi^{\ast} : \E \rightarrow \E$, given by $\phi^{\ast}(a) = a \circ \phi$ for any $a \in \E$. Operationally, to measure $\phi^{\ast}(a)$ on a state $\alpha$, one first subjects the state $\alpha$ to the process $\phi$, and then makes a measurement of the effect $a$. Note that $\tau(u)(\alpha) = u(\tau^{\ast}(\alpha))$ is 
the probability that the process $\tau^{\ast} = \phi$ occurs if the initial state is $\alpha$. In what follows, it will 
generally be more convenient to deal with these dual processes; accordingly, we'll broaden our usage and refer to these, 
also, as processes.

{\bf Probabilistic Models and Theories} In view of the preceding discussion, the following language seems reasonable. 

\begin{definition} {\em A finite-dimensional {\em probabilistic model} is a triple $A = (\E(A),\A(A),\D(A))$ consisting of 
\begin{itemize}
\item[(i)] a finite-dimensional order-unit space $(\E(A),u_{A})$, 
\item[(ii)] a test 
space $\A$ consisting of observables on $\E(A)$, with outcome-set $X =
\bigcup \A$ generating $\E_+(A)$, and 
\item[(iii)] a semigroup $\D(A)$ of
positive mappings 
$\tau : \E \rightarrow \E$, called {\em processes}, satisfying
$\tau(u) \leq u$. 
\end{itemize}
A {\em state} of the model is a normalized, positive
linear functional $\alpha : \E(A) \rightarrow \R$.} \end{definition}


Broadly speaking, a {\em probabilistic theory} is a class $\cc$ of
such models. In particular, we can identify finite-dimensional quantum
theory with the class of models in which $\E$ is the set of hermitian
elements of a complex matrix algebra ${\cal A}$, with the usual
operator-theoretic ordering, $u$ is the identity functional, $\A$
consists of maximal, pairwise orthogonal sets of projection operators,
and $\D$ is the semigroup of completely
positive maps on $\cal A$.

{\bf Reversible Processes} We shall say that a physical process
$\phi$, or the dual process $\tau = \phi^{\ast}$, is {\em physically
reversible} iff it is invertible as a linear mapping, with a positive
inverse --- that is, $\phi$ is an {\em order-automorphism} of
$\E(A)^{\ast}$ --- and $\phi^{-1}$ is a positive multiple of a
physical process --- say, $\phi^{-1} = c \phi_o$ for some process
$\phi_o$.  Operationally, this means that there is always some
non-zero probability that $\phi_o \circ \phi$ will return the system
to its original state.  Indeed, for any normalized state $\alpha$, 
\[\phi_o(\phi(\alpha))(u) = \phi_o(c \phi_{o}^{-1}(\alpha))(u) = c \alpha(u) = c,\]
so this probability --- which is independent of the initial state $\alpha$ --- is exactly the factor $c$. 
Notice that $\phi$ is reversible with probability one iff $c = 1$, i.e., $\phi^{-1}$ is a
process.\footnote{Many authors 
define ``reversible'' by this condition, i.e. as what we have here 
called reversible with probability one. }  This implies that $\tau = \phi^{\ast}$ satisfies $\tau(u) =
u$. Conversely, if $\tau = \phi^{\ast}$ and $\tau u = u$, then
$\tau^{-1} u = u$. Thus, if $\tau^{-1} = c \tau_o$, where $\tau_o$ is
a process, then, on states, then the probability of
\[(c \tau_{o}^{\ast})(\alpha)(u) = c \alpha(\tau(u)) = c \alpha(u) = c.\]
Clearly, the set $\D_1(A)$ of invertible processes forms a
sub-semigroup of $\D(A)$, and generates a subgroup, $\G(A)$, of
$\Aut(\E(A))$, namely, the set of all multiples $c\tau$ where $\tau \in
\D_1$ and $c \in {\mathbb R}_+$. 
Those processes reversible with
probability $1$ are exactly the invertible processes $\tau \in \D(A)$
with $\tau(u) = u$, i.e., those in the stabilizer $\G(A)_{u_{A}}$.

\subsection{The Jordan structure of an HSD model}

Our proof of Proposition 1, given in Section 3, depends on the details of the construction of the Jordan
product on an HSD order-unit space. 
In what follows, let $(\E,u)$ be an HSD order-unit
space. By this we mean a finite-dimensional order-unit space $\E$, the
positive cone of which is homogeneous, and for which there {\em
exists} an inner product making $\E_{+} = \E^{+}$. We call such an
inner product {\em self-dualizing}.\footnote{This differs slightly, but 
not materially, from the definition of an HSD cone in
\cite{Faraut-Koranyi}, where a fixed inner product is assumed.}
Let $G$ be any closed subgroup of $\Aut(\E)$, acting transitively on
the interior of $\E_+$.  Then $G$ is a Lie subgroup of $GL(\E)$. Let
$\g$ denote its Lie algebra, and let $\g_u$ denote the Lie algebra of
the stabilizer $G_u \leq G$ of the order-unit. 
The following formulation 
of the Koecher-Vinberg Theorem summarizes the construction of the Jordan product on $\E$.

\begin{theorem}[Koecher-Vinberg] Let $G$ be a closed, connected subgroup of $\Aut(\E)$, acting transitively on the interior of $\E_{+}$. Then 
\begin{itemize} 
\item[(a)] It is possible to choose a self-dualizing inner product on
$\E_+$ in such a way that $G_u = G \cap \O(\E)$ (where $\O(\E)$ is the 
orthogonal group with respect to the inner product);
\item[(b)] If $G = G^{\dagger}$ with respect to this inner product, then
$\g_u = \{ X \in \g | X^{\dagger} = -X\} = \{ X \in \g | Xu = 0\}$, and $\g = \g_u \oplus \p$, where
$\p = \{ X \in \g | X^{\dagger} = X\}$;
\item[(c)] In this case the mapping $\p \rightarrow \E$, given by $X
\mapsto Xu$, is an isomorphism.  Letting $L_a$ be the unique element
of $\p$ with $L_a u = a$, define
\[a \bullet b = L_{a} b\] 
for all $a, b \in \E$. Then $\bullet$ makes $\E$ a formally real
Jordan algebra, with identity element $u$.
\end{itemize} \end{theorem}

{\em Remark:} The proof of the Koecher-Vinberg Theorem given in
\cite{Faraut-Koranyi} takes $G$ to be the connected identity component
of the automorphism group of $\E$. We are making the ostensibly stronger 
claim here that any homogeneously-acting, closed, self-adjoint
subgroup of $\Aut(\E)$ will suffice; accordingly, a detailed sketch of
the proof is given in an Appendix to this paper.

\subsection{HSD and Jordan models} 

We shall say that a model $A$ is {\em HSD} (homogeneous and self-dual)
iff the cone $\E_+(A)$ is homogeneous under its group $\G(A)$ of
reversible processes, and equal to its dual with respect to {\em some}
inner product.  If $A$ is an HSD model, then the Koecher-Vinberg
theorem implies that $\E(A)$ carries a unique Euclidean Jordan
structure with respect to which the order unit, $u_A$, is the
identity.

An {\em idempotent} in $\E(A)$ is a non-zero element $p \in \E_+(A)$
such that $p^2 = p$ (where $p^2 = p \bullet p$). A non-zero
idempotent that cannot be decomposed as the sum of two distinct
non-zero idempotents is said to be {\em primitive}. The spectral
theorem for Euclidean Jordan algebras (see \cite{Faraut-Koranyi},
Proposition III.1.2) tells us that every nonzero element of $\E_+(A)$
is the sum of positive multiples of pairwise-orthogonal primitive
idempotents. It follows that every extremal ray of $\E(A)_+$ consists
precisely of the nonnegative multiples of some primitive idempotent,
idempotent generates such an extremal ray. Since the set $X(A)$ of
outcomes of the model $A$ generates the positive cone $\E_+(A)$, we
can conclude that every primitive idempotent is a positive multiple of
some outcome.  However, $X(A)$ may also contain some non-extremal
outcomes.  In this section, we identify two simple and natural
conditions that together guarantee that every outcome {\em is}, in
fact, a primitive idempotent.

For the balance of this section,  $A$ is an HSD model, equipped with its 
corresponding Jordan structure and trace, and with the tracial inner product defined by 
$\langle a, b \rangle = \tr(a b)$ for all $a,b \in \E(A)$. Notice that $\langle a, b \rangle \geq
0$ for all $a, b \in \E(A)_+$. A primitive idempotent
$e \in \E(A)$ satisfies $\tr(e) = 1$; hence, by the
Cauchy-Schwarz inequality, $\langle e, f \rangle \leq 1$ for
all primitive idempotents $f$. We also have $\langle e, e \rangle = \langle e, u \rangle = \tr(e) = 1$. 
Thus, a primitive idempotent $e$ defines a pure state,
$\langle e |$ on $A$, and this is the unique pure state assigning probability
$1$ to the effect corresponding to $e$.

A {\em Jordan frame} in a Euclidean Jordan algebra $\E$ is a set
$e_1,...,e_n$ of primitive idempotents summing to $u$.  
All Jordan
frames in $\E$ have the same cardinality, called the {\em rank}
of $\E$.  By a {\em Jordan model}, we mean an HSD model such that
every outcome is a primitive idempotent, or, equivalently, every test
is a Jordan frame.

Let us say that a probabilistic model $A$ is {\em uniform} iff there
exists a state $\mu \in \E(A)^{\ast}$ taking a constant value $\mu(x)
= 1/m$ on all outcomes $x \in X(A)$. Note that this implies that all
tests $E \in \A(A)$ have cardinality $m$. An outcome $x \in X(A)$ is
{\em unital} iff  there exists a state $\alpha \in \E^{\ast}$ with
$\alpha(x) = 1$, and {\em sharp} if this state is unique.  The model
$A$ itself is unital, respectively, sharp, iff every outcome $x \in
X(A)$ is unital, respectively, sharp.  Observe that any Jordan model
is sharp (hence, unital) and uniform, with uniform state given by
$\mu(x) = \langle u, x \rangle = 1/n$, $n$ the rank of $\E$. We now
establish the converse.

\begin{lemma} Let $A$ be HSD. 
\begin{itemize} 
\item[(a)] Every extremal unital outcome is a primitive idempotent. 
\item[(b)] If $A$ is uniform, then every unital outcome is extremal, hence, a primitive idempotent.
\end{itemize} \end{lemma}

{\em Proof:} (a) Let $x \in X(A)$ be extremal. As observed above, there exists some $t > 0$ such that $tx =: e$, a primitive 
idempotent. Now suppose $f$ is a primitive idempotent representing a pure state of $\E$, with $\langle f, x \rangle = 1$. 
Then 
\[t = t \langle f, x \rangle = \langle f, tx \rangle = \langle f, e \rangle \leq 1,\]
by the Cauchy-Schwarz inequality. Now notice that 
\[t^2 \langle x, x \rangle = \langle e, e \rangle = 1\]
so $\langle x, x \rangle = 1/t^2$. Choosing any $E \in \A(A)$ with $x \in E$, we now have 
\begin{eqnarray*}
1 = \langle e, u \rangle & = & t \langle x, u \rangle \\
& = & t ( \langle x, x \rangle + \sum_{y \in E \setminus \{x\}} \langle x, y \rangle\\
& \geq & t \langle x, x \rangle =  t/t^2 = 1/t,\end{eqnarray*}
so that $t \geq 1$. Thus, $t = 1$, and $x = e$. 

(b) Let $x \in X(A)$ and 
$x = \sum_{i} s_i x_i$ where the $x_i$ are extremal outcomes and $s_i \geq 0$. 
Let $\mu$ be the uniform state on $\E$. Then 
\[\frac{1}{m} = \mu(x) = \sum_{i} s_i \mu(x_i) = \sum_i s_i \frac{1}{m} \]
so $\sum_i s_i = 1$. If $x$ is unital, therefore, there exists a pure state assigning probability $1$ to $x$; hence, by 
the self-duality of $\E(A)_+$, there exists a primitive idempotent $f$ with 
\[1 = \langle f, x \rangle = \sum_{i} s_i \langle f, x_i\rangle.\]
Since, as we've just seen, $s_i \ge 0$ and $\sum_i s_i = 1$, we have
$\langle f, x_i \rangle = 1$ for every $i$ with $s_i \not = 0$. But
then, every $x_i$ is a unital extremal outcome and so, by part (a), a
primitive idempotent. It follows (again by Cauchy-Schwarz and the
argument in the proof of (a)) that $s_i \not = 0$ implies $x_i = f$,
whence, $x = f$. $\Box$

It follows that any HSD model that is both uniform and unital is a
Jordan model. Since, as observed above, the converse also holds,
uniform, unital HSD models are exactly the same things as Jordan
models. Notice that any Euclidean Jordan algebra $\E$ can be equipped
with the structure of a Jordan model by choosing a distinguished
family $\A$ of Jordan frames such that the set $X = \bigcup \A$
generates $\E_+$.  In particular, we can always take $\A$ to be the
set of {\em all} Jordan frames.

\section{Composites of Jordan Models} 

We now wish to examine the structure of composite systems comprising
two Jordan models. We begin with a review of the notion of a composite
of probabilistic models, following \cite{BBLW07, BW09b}.

\subsection{Composites and tensor products} 

Consider two systems $A$ and $B$, which, while possibly interacting,
retain enough independence to allow them to be observed and
manipulated separately.  We would then expect a model for the
composite system $AB$ to include, for each pair of effects $a \in
\E(A)$, $b \in \E(B)$, a {\em product effect} $a \otimes b \in
\E(AB)$, with the understanding that, for a state $\omega \in
\E(AB)^{\ast}$, $\omega(a \otimes b)$ gives the joint probability to
observe $a$ and $b$. Moreover, we should expect that the two systems
can be prepared independently in arbitrary states $\alpha \in
\E(A)^{\ast}$, $\beta \in \V(B)$, so as to produce a {\em product
state} $\alpha \otimes \beta$ with $(\alpha \otimes \beta)(a \otimes
b) = \alpha(a)\beta(b)$. Finally, if $g_A \in G(A)$ and $g_B \in G(B)$
are symmetries of $A$ and $B$, respectively, then there should exist a
symmetry $g \in G(AB)$ such that $g(a \otimes b) = ga \otimes gb$ for
all $a \in \E(A)$ and $b \in \E(B)$.

Supposing this much, let $\omega$ be a state on $\E(AB)$. We shall say
that $\omega$ is {\em non-signaling} iff the two {\em marginal states}
given by
  \[\omega_{A}(a) := \sum_{y \in F} \omega(a \otimes y) \ \ \text{and} \ \ \omega_{B}(b) := \sum_{x \in E}
  \omega(x \otimes b)\] are well-defined, i.e., independent of the
  choice of tests $E \in \A(A)$ and $F \in \A(B)$ (This prevents
  parties controlling $A$ and $B$ from sending one another information
  solely by choosing which tests to measure.)  It is not hard to see
  \cite{Wilce92, BFRW} that this makes the mapping $a,b \mapsto
  \omega(a,b)$ bilinear, whence, $\alpha, \beta \mapsto \alpha\otimes
  \beta$ and $a, b \mapsto a \otimes b$ are also bilinear, justifying
  the tensorial notation.  These considerations motivate the following
  definition.

\begin{definition} \label{def: composite}
A {\em non-signaling composite} of (finite-dimensional) models $A$ and
$B$ is a model $AB$, equipped with two bilinear mappings
\[\otimes : \E(A)
\times \E(B) \rightarrow \E(AB) \ \text{and} \ \otimes : \E(A)^{\ast} \times \E(B)^{\ast} \rightarrow
\E(AB)\] such that
\begin{itemize} 
\item[(i)] For all tests $E \in \A(A)$ and $F \in X(B)$,  $E \otimes F = \{ x \otimes y | x \in E, y \in F\}$ is 
a test in $\A(AB)$; 
\item[(ii)] $(x \otimes y)(\alpha \otimes \beta) = \alpha(x)\beta(y)$ for all states 
$\alpha \in \E(A)^{\ast}, \beta \in \E(B)^{\ast}$;
\item[(iii)] For all $\tau_A \in \D(A)$ and $\tau_B \in \D(B)$, there exists a process $\tau \in \G(AB)$ such that 
\[\tau(\alpha \otimes \beta)  = \tau_A \alpha \otimes \tau_B \beta\]
for all $\alpha \in \E(A)^{\ast}, \beta \in \E(B)^{\ast}$.   
\end{itemize} 
\end{definition}

It follows from (i) that if $x$ and $y$ are outcomes of $A$ and $B$, then $x \otimes y$ is an outcome of $AB$. This, 
together with condition (ii) and the bilinearity of the mappings $\otimes$, that $(\alpha \otimes \beta)(a \otimes b) = \alpha(a) \beta(b)$ for all $a \in \E(A), b \in \E(B)$ and all $\alpha \in \E(A)^{\ast}, \beta \in \E(B)^{\ast}$. We also 
have $u_{A} \otimes u_{B} = u_{AB}$. 

\begin{definition}
We say that $AB$ is {\em locally tomographic} iff
every bipartite state $\omega \in \E(AB)^{\ast}_{+}$ is entirely determined by
the joint probabilities $\omega(a,b) :=  \omega(a \otimes b)$ that
$\omega$ assigns to pairs of effects $a \in \E(A), b \in \E(B)$.
\end{definition} 

If $A$ and $B$ are finite-dimensional (that is, if $\E(A)$ and $\E(B)$ are finite-dimensional), the 
condition that a non-signaling composite be local tomographic 
is equivalent to the condition
that $\dim(\E(AB)^{\ast}) = \dim(\E(A)^{\ast})\dim(\E(B)^{\ast})$, that, is, as vector spaces (ignoring the order structure) $\E(AB)^{\ast} = \E(B)^{\ast} \otimes \E(B)^{\ast}$ and $\E(AB) = 
\E(A) \otimes \E(B)$.  Note that this makes the process $\tau \in \D(AB)$
required by condition (iii) above unique, so that we can sensibly
write $\tau = \tau_A \otimes \tau_B$.

\subsection{Proof of Theorem 1} 

We now consider the implications of the existence of a locally
tomographic HSD composite, $AB$, of HSD systems $A$ and $B$.
Recalling the notation used in Section 3, if $A$ is any system, let
$G(A)$ denote the connected identity component of the group $\G(A)$ of
invertible physical processes on $A$.  If $\G(A)$ acts homogeneously
on $\E_+$, so does $G(A)$ (\cite{Faraut-Koranyi}, p. 5).

Since $AB$ is HSD, we can introduce an inner product $\langle , \rangle_{AB}$ on $\E(AB)$ that is normalized 
and self-dualizing for $\E(AB)_+$. 
We shall say that $\langle , \rangle_{AB}$ {\em factors}, and that $\E(AB)$ is {\em factorizably self-dual}, iff 
\[\langle a \otimes b, c \otimes d \rangle = \langle a, c \rangle \langle b, d \rangle\]
for all $a, c \in \E(A)$ and $b, d \in \E(B)$ where $\langle,
\rangle_{A}$ and $\langle , \rangle_{B}$ are self-dualizing inner
products on $\E(A)$ and $\E(B)$, respectively.  More generally, say
that $AB$ is {\em factorizably self-dual} iff the self-dualizing inner
product can be chosen to factor in this way. This is not as
restrictive a condition as it might at first seem:

\begin{lemma} Let $AB$ be a non-signaling composite of Jordan models $A$ and $B$. If $AB$ is itself Jordan, 
then the trace form on $\E(AB)$ factors. \end{lemma} {\em Proof:} By
the definition of a composite, if $x, y \in X(A)$, then $x \otimes y$
is an outcome in $X(AB)$. Since $x$ and $y$ are unital in $A$ and $B$, $x
\otimes y$ is unital in $X(AB)$: the product state $\langle x |
\otimes \langle y|$ assigns $x \otimes y$ probability $1$ (again, by
the definition of a composite). Hence, by Lemma 6 (b), $x \otimes y$
is a primitive idempotent in $\E(AB)$, and therefore pure.  But 
since $AB_+$ is HSD, there is a unique pure
state, $\langle x \otimes y |$, with $\langle x \otimes y | x
\otimes y \rangle = 1$.  Hence, $\langle x |
\otimes \langle y | = \langle x \otimes y |$, so that 
$\langle x \otimes y | a \otimes b \rangle = \langle x | a \rangle \langle y | b \rangle$ 
for all $a\in \E(A), b \in \E(B)$. Since $X(A)$ spans $\E(A)$ and
$X(B)$ spans $\E(B)$ the same holds with arbitrary elements of $\E(A)$
and $\E(B)$ in place of $x$ and $y$ respectively, i.e, the inner
product factors. $\Box$

{\em Remark:} Rather than assuming that $AB$ is Jordan (equivalently,
HSD, uniform, and unital), one can suppose that it is HSD
and that every $E \in \A(AB)$ has the same cardinality. Since
$\A(AB)$ contains product tests, the cardinality of its tests must then be
$mn$ where $m$ and $n$ are the ranks of the Jordan algebras $\E(A)$ and $\E(B)$.
Now the product of the uniform state on $A$ with the uniform state on
$B$ provides a uniform state on $AB$. Lemma 6 can then be invoked, as in 
the proof above.

\begin{lemma} Let $AB$ be locally tomographic and factorizably self-dual.  
\begin{itemize} 
\item[(a)] If $g \in GL(\E)$, then $(g \otimes \1_B)^{\dagger} = g^{\dagger} \otimes \1_B$
\item[(b)] If $g \in G(A)$, then $g^{\dagger} \otimes \1_B$ is an order-automorphism 
of $\E(AB)$. 
\end{itemize} \end{lemma}

{\em Proof:} (a) For elements of $\E(AB)$ of the form $a
\otimes b$,
we have
\[\langle (g \otimes \1_B) a_1 \otimes b_1 , a_2 \otimes b_2 \rangle = 
\langle g a_1, a_2 \rangle \langle b_1, b_2 \rangle = \langle a_1, g^{\dagger} a_2 \rangle \langle b_1, b_2 \rangle = 
\langle a_1 \otimes b_1, (g^{\dagger} \otimes \1_B) a_2 \otimes b_2 \rangle.\]
Since $AB$ is locally tomographic, such elements span $\E(AB)$, so the
relation holds generally.

(b) This now follows, since the adjoint of an order-automorphism with respect to a self-dualizing inner product is again 
an order-automorphism (cf. \cite{Faraut-Koranyi}, I.1.7). $\Box$

Let $AB$ be a locally tomographic, HSD composite of HSD models $A$ and
$B$, and suppose $\langle ~, ~ \rangle_{AB}$ is a factorizable
self-dualizing inner product on $\E(AB)$, where the inner products on
the factors are chosen in accordance with Theorem 8 (a). Let 
$G(A)^{\dagger}$ denote the group of adjoints $g^{\dagger}$ where $g \in G(A)$, 
with the adjoint defined by the chosen inner product. As noted above, the self-duality 
of $\E(A)$ ensures that $G(A)^{\dagger} \leq \Aut(\E)$. Now let 
\[G_{A} := \langle G(A) \cup G(A)^{\dagger} \rangle,\]
that is, $G_A$ is the closed subgroup 
of $\Aut(\E)$ generated by $G(A)$ and $G(A)^{\dagger}$. The group $G_{A}$ 
acts transitively on the interior of $\E_{+}$, since $G(A)$ does, and is easily seen to be
self-adjoint and connected. 
Theorem 8 therefore yields, for each $a \in \E(A)$, a unique
self-adjoint element $L_a$ of the Lie algebra $\g_A$ of $G_A$, such that $L_a u_{A} = a$, and such
that $a \bullet b := L_{a} b$ defines the Jordan structure of $\E(A)$. 
Moreover, we have 

\begin{lemma} If $g \in G_{A}$, then $g \otimes \1_B \in G_{AB}$. \end{lemma} 

{\em Proof:} For every element $g \in G(A)$, $g \otimes \1_B \in G(AB)$ (by condition (iii) of Definition 9), 
whence, by Lemma 10, $g^{\dagger} \otimes \1_B$ 
belong to $G(AB)^{\dagger}$. $\Box$ 

Thus, we have a canonical embedding $G_{A} \simeq G_{A} \otimes \{\1_B\} \leq G_{AB}$. 
As in Theorem 4, let $\g_A$ denotes the Lie algebra of $G_{A}$ and $\p_{A}$, the self-adjoint part of $G_{A}$, 
and similarly for $\g_{AB}$ and $\p_{AB}$.  

\begin{lemma} 
 For all $a \in \E(A), v \in \E(B)$, $L_a \otimes \1_B, \1_A \otimes L_v
\in \g_{AB}$. Hence, $L_{a \otimes u_B} = L_{a} \otimes \1$ and
$L_{u_A \otimes v} = \1_A \otimes L_{v}$.
\end{lemma}

\noindent{\em Proof:}  Since $G_A \leq G_{AB}$ (via $g
\mapsto g \otimes \1_B)$, 
and as $L_a \in \p_{A}$,  
we have a one-parameter group $g : t
\mapsto g_t \in G_{A}$ with $L_a = g'(0)$, and a corresponding
one-parameter group $g \otimes \1 : t \mapsto g_t \otimes \1$ in
$G_{AB}$. The bilinearity of the tensor product  
 gives us $L_a \otimes \1 = (g \otimes \1)'(0) \in
\g_{AB}$.  Since the inner product factors, we have 
\[(L_a \otimes \1)^{\dagger} = (L_a)^{\dagger} \otimes
\1^{\dagger} = L_a \otimes \1\] 
So $L_x \otimes \1$ is a self-adjoint element of $\g_{AB}$, 
that is, an element of $\p_{AB}$. Also, $(L_a \otimes \1)(u_A \otimes
u_B) = L_a u_A \otimes u_B = a \otimes u_B$.  The second identity is proved similarly. $\Box$

\begin{corollary} \label{cor: main} In $AB$, we have 
\[(a \otimes u_{B})\bullet (b \otimes v) = (a\bullet b) \otimes v \ \ \text{and} \ \ (u_{A} \otimes v)\bullet (a \otimes w) = 
a \otimes (v \bullet w).\]\end{corollary} 

\noindent{\em Proof:} Taking the first identity, we have 
\[(a \otimes u_{A})\bullet (b \otimes v) = L_{a \otimes u_B}(b \otimes v) = (L_a \otimes \1_B)(b \otimes v) =  L_a b 
\otimes v = (a \bullet b) \otimes v.\]
Similarly for the second identity. $\Box$ \\

This corollary is what we promised to establish, namely that a locally
tomographic, factorizably HSD composite of HSD models must satisfy
Hanche-Olsen's condition, (\ref{eq: Hanche-Olsen condition}) --- hence
, by Hanche-Olsen's {\em theorem}, the models involved are
self-adjoint parts of complex $C^{\ast}$-algebras, giving Proposition
1.\footnote{The reader
  may have noticed that the only explicit use we've made of local
  tomography in the derivation of Corollary \ref{cor: main} is in the
  proof of part (a) of Lemma 11. However, to apply Hanche-Olsen's
  theorem together with Corollary \ref{cor: main}, we need to know that
  the ordinary vector space tensor product $\E(A) \otimes \E(B)$,
  where $B$ is a qubit, has an HSD structure; this only follows if the composite
  $AB$ is locally tomographic.} In the language of Section 2.2, and
appealing to Lemma 10, we can rephrase this as asserting that {\em any
  locally tomographic, non-signaling theory in which all models are
  Jordan models, is a standard quantum theory.}

In the next section, we consider the implications of this result in the setting of a dagger-monoidal category $\cc$ of HSD probabilistic models.

\section{Categorical Considerations}\label{sec: categories}

It is reasonable
to represent a physical
theory as a category, $\cc$, in which objects represent distinct
physical systems and morphisms represent physical processes.  To
capture the idea that processes can be composed not only serially, but
also in parallel, it's equally natural to supose that $\cc$ carries a
symmetric monoidal structure.  This point of view has been developed
extensively by Abramsky and Coecke \cite{Abramsky-Coecke}, Baez
\cite{Baez}, and Selinger \cite{Selinger}. A striking result of this
work is that many qualitative features of quantum information
processing are actually direct consequences of the fact that the
category of finite-dimensional Hilbert spaces and linear mappings is a
{\em dagger-compact} category.

We recall that a {\em symmetric monoidal category} is a category
$\cc$, equipped with a bi-functorial operation $\otimes: \cc \times \cc
\rightarrow \cc$ that is commutative and associative, up to natural
isomorphisms $A \otimes B \simeq B \otimes A$ and $A \otimes (B
\otimes C) \simeq (A \otimes B) \otimes C$ for objects $A, B, C$ of
$\cc$, and also equipped with a {\em tensor unit}, $I$, such that $I
\otimes A \simeq A$ for all $A \in \cc$.  We shall be interested here
in symmetric monoidal categories having probabilistic models as objects. 
It would be natural to take morphisms to be those positive linear mappings that 
we wish to consider physical processes; for computational convenience,
however, it seems reasonable to enlarge the set of morphisms to
include arbitrary linear combinations of such processes.  This suggests the
following:

\begin{definition}{\em A finite-dimensional {\em  monoidal probabilistic theory} is a symmetric monoidal category $\cc$ in which 
\begin{itemize} 
\item[(i)] Objects are finite-dimensional probabilistic models;
\item[(ii)] For all objects $A, B \in \C$, the set $\C(A,B)$ of morphisms $A \rightarrow B$ is a
non-trivial subspace of the space $\cl(\E(A),\E(B))$ of linear mappings
$\phi : \E(A) \rightarrow \E(B)$, equipped with a generating cone
$\C_{+}(A,B)$ of positive linear mappings.  Composition of morphisms
is composition of mappings, and $\C_+(A,B) \circ \C_+(B,C) \subseteq
\C_+(A,C)$ for all $A,B,C \in \C$; 
\item[(iii)] The tensor unit $I$ is ${\mathbb R}$ (with its usual order
and unit), and $\C(I,A) = \E(A)$.
\item[(iv)] For every $A \in \C$, $\D(A)$ is the set of all morphisms
$\phi \in \C_{+}(A,A)$ satisfying $\phi(u_A) \leq u_B$.  
Accordingly, the group $G(A)$ generated by reversible processes in $\D(A)$, is exactly the group of invertible elements of
$\C(A,A)$ having inverses in $\C_{+}(A,A)$. 
\item[(v)] The monoidal product, $AB$, of two objects $A, B \in \C$,
  is a locally tomographic composite of the models $A$ and $B$, in the
  sense of Definition 8, and the monoidal operation $\C(A,A')
  \times \C(B,B') \rightarrow \C(AA',BB')$ is bilinear for all $A',B'
  \in \C$.
\end{itemize}  
}\end{definition}

By condition (ii), $\C(A,I)$ is a non-trivial subspace of $\E(A)^{\ast}$, but, absent
some further constraint, these need not be isomorphic. Indeed, the requirement 
that $\C(A,I) \simeq \E(A)^{\ast}$ is equivalent to requiring that every state $\alpha$ on $A \in \C$ 
correspond to a morphism $\alpha : A \rightarrow I$, hence, to an element of $\E(A)^{\ast}$. This is precisely 
the state-completeness assumption discussed in Section 2. Given state-completeness, if $\alpha \in \E(A)^{\ast} \simeq \C(A,I)$ and $a \in \E(A) \simeq \C(I,A)$, we have 
\[\alpha(a) = \alpha \circ a.\] 

\begin{lemma} Let $\C$ be a state-complete  monoidal probabilistic theory. Then for every pair of objects 
$A,B \in \C$, the composite $AB$ is non-signaling.\end{lemma}

{\em Proof:} 
If $\C$ is state-complete, the linearity of morphisms and 
the bilinearity of $\otimes$ together imply that composite systems in $\C$ are non-signaling. 
Suppose $\omega : A \rightarrow I$. If $E \in \A(A)$ and $y \in X(B)$, then --- identifying each $x \in E$ with the corresponding linear 
mapping $x : I \rightarrow A$, and similarly for $y$ --- we have 
\[\sum_{x \in E} \omega(x,y) := \sum_{x \in E} \omega \circ (x \otimes y) = \omega(\sum_{x \in E} x \otimes y) = 
\omega(u_A,y)\]
Similarly, $\sum_{y \in F} \omega(x,y) = \omega(x,u_B)$ for all $F \in \A(B)$. $\Box$ 

From this point on, we assume that $\C$ is state-complete.

A {\em dagger} \cite{Selinger} on a category $\cc$ is a contravariant
endo-functor\footnote{A contravariant endo-functor $F$ on a category
  $\cc$ is a map from the morphisms of $\cc$ to the morphisms of
  $\cc$, such that $F(f \circ g) = F(g) \circ F(f)$ (as opposed to an
  ordinary (covariant) endo-functor, for which $F(f \circ g) = F(f)
  \circ F(g)$).}  $\dagger : \cc \rightarrow \cc$ such that, for all
objects $A \in \cc$, $A^{\dagger} = A$, and, for all morphisms $\phi,
\psi$ in $\cc$, $\phi^{\dagger\dagger} = \phi$.
An morphism $\phi$ in $\cc$ is {\em unitary} with respect to $\dagger$ iff 
it is invertible and satisfies $\phi^{-1} = \phi^{\dagger}$. A {\em dagger-monoidal
category} is a symmetric monoidal category equipped with a dagger
that commutes with the
monoidal structure, so that $(\phi \otimes \psi)^{\dagger} = \phi^{\dagger}
\otimes \psi^{\dagger}$ for all morphisms $\phi$ and $\psi$ in $\C$, and is 
such that the isomorphisms comprising the
components of the natural associativity, symmetry, and
unit-introduction transformations are unitary.

Let $\cc$ be a monoidal probabilistic theory, equipped with a
dagger. Assume, further, that the dagger operation is linear,
and positive; that is, if $\phi \in \C_+(A,B)$, then $\phi^{\dagger}
\in \C_+(B,A)$.  In this setting, we have $\C(A,I) \simeq \C(I,A) = \E(A)$, whence, by dimensional
considerations, $\C(A,I) \simeq \E(A)^{\ast}$ as a linear space.
\footnote{Note that we take the states to be elements of $\C(A,I)$,
while the effects belong to $\C(I,A)$.  This is the reverse of the
convention in many papers (e.g. \cite{Abramsky-Coecke, Selinger}, and
ourselves in \cite{BDW10}), and may be thought of as working in the
``generalized Heisenberg picture'' for probabilistic theories.  It is
motivated in part by the naturality of viewing states as functionals
from an order-unit space (or test space) to $\R$, and effects as dual
to these.}  also have $\C_{+}(A,I)
\simeq \C(I,A)$, but the possibility remains open that $\C_+(A,I)$ is
a proper sub-cone of the dual cone $\E(A)^{\ast}_+$.

The dagger also provides us with a canonical bilinear form on each $\E(A)$, $A \in \C$, defined by  
\[\langle a, b \rangle_A := b^{\dagger} \circ a\]
for all $a, b \in \cc(I,A) \simeq \E(A)$.
Since $r^{\dagger} = r$ for every $r \in \R = I$, this form is symmetric: 
\[\langle b, a \rangle = a^{\dagger} \circ b = (b^{\dagger} \circ a)^{\dagger} = \langle a, b \rangle^{\dagger} = \langle a, b \rangle.\]
 An element $g$ of the group $\G(A)$ of invertible morphisms in $\C(A,A)$ is said to be {\em unitary} iff $g^{\dagger} = g^{-1}$. We write $U(A)$ for the group of unitary elements of $\G(A)$. The bilinear form $\langle , \rangle_A$ is invariant with respect to $U(A)$, in the sense that 
\[\langle g a, g b \rangle_A = (g \circ b)^{\dagger} \circ g \circ a = b^{\dagger} \circ g^{\dagger} \circ g \circ a 
= b^{\dagger} \circ a = \langle a, b \rangle_A.\] This bilinear form
is also {\em positive}, in the sense that $\langle a, b \rangle \geq 0$
for $a, b \in A_+$.\footnote{This is quite distinct from positive-definiteness, 
i.e., we are not claiming that $\langle a, a
\rangle > 0$ for all $a \ne 0$.}  Finally, observe (from the bifunctoriality of
$\otimes$) that the canonical bilinear form {\em factors}, in the
sense that, for every $A, B \in \C$,
\begin{equation} \langle a \otimes b, c \otimes d \rangle_{AB} = \langle a, c \rangle_{A} \langle b, d \rangle_{B}\end{equation}
for all $a, c \in \C(I,A) \simeq \E(A)$ and all $c, d \in \C(I,B)
\simeq \E(B)$.

\begin{definition} {\em A {\em dagger-monoidal probabilistic theory} is a 
(state-complete) monoidal probabilistic theory $\cc$ equipped with a
    (positive, linear) dagger, such that, for every $A \in \cc$, and
    every $g \in U(A)$, $g u_A = u_A$. We shall call $\cc$ {\em
      dagger-HSD} iff (i) the canonical form $\langle , \rangle_A$ is
    an inner product (i.e., is positive semidefinite)
 and (ii) every system
$A$ in $\cc$ is homogeneous with respect to $\cc(A,A)$. }
\end{definition}

{\em Remark:} The condition that $\langle , \rangle$ be an inner product is
non-trivial, but can be motivated in the case in which every system $A \in \cc$ is 
an irreducible Jordan model. In this case, using \cite{Wilce11}, Corollary 2 and Lemma 6, 
one can show that any positive, $G(AA)$-invariant bilinear form on $AA$ will in fact be an 
inner product. 

Let $\C$ be any dagger-HSD probabilistic theory. It is straightforward to verify that, if $A \in \C$ 
and $\tau \in \C(A,A)$, then
$\tau^{\dagger} \in \C(A,A)$ functions as the adjoint of $a$ with
respect to the canonical inner product. That is, if $a, b \in \C(I,A)
\simeq A$, then
\[\langle \tau a, b \rangle = (\tau \circ a)^{\dagger} \circ b = a \circ \tau^{\dagger} \circ b = \langle a, \tau^{\dagger} b \rangle.\]
In particular, then, the group $\G(A)$ is self-adjoint with respect to
$\langle, \rangle_A$ (i.e. $x \in G(A) \implies x^\dagger \in
G(A)$). It follows that the cone $\E(A)_+$ is self-dual
(cf. \cite{Faraut-Koranyi}, Exercise I.8).

In view of Proposition 1, the factorization property (\theequation)
immediately yields the corollary that every locally tomographic
dagger-HSD theory containing at least one qubit, is a standard quantum
theory in the sense that all of its systems are isomorphic, as ordered
linear spaces, to the self-adjoint parts of complex matrix algebras
with their standard orderings, and the monoidal product is the usual tensor
product of quantum systems. That is,

\begin{proposition}
A  locally tomographic, dagger-HSD probabilistic theory in which at
least one system has the structure of a qubit, is a standard quantum
theory. \end{proposition}

\section{Conclusions} 

We have shown that, in the specific context of (finite-dimensional, state-complete) 
Jordan probabilistic models --- that is, uniform, unital models with homogeneous, self-dual
cones ---  defining features of orthodox, complex QM are
\begin{itemize}
\item[(i)] the availability of locally tomographic, non-signaling
products --- otherwise, a weak constraint;
\item[(ii)] the existence of a qubit. 
\end{itemize}
Similarly, in the context of a state-complete dagger-HSD theory (where
composites are automatically non-signaling), quantum theory is picked
out by local tomography and the existence of a quibit.

In \cite{Dakic-Brukner}, Daki{\'c} and Brukner derive QM from assumptions
that also include (in effect) the existence of a qubit; however, they
make a very strong uniformity assumption, namely, that all systems of a
given information-carrying capacity, are isomorphic.  This is not
unreasonable if we imagine that all systems are built up, through a
uniform process of composition, from a single elementary system --- in
this case, a qubit. And, indeed, in \cite{Muller2}, it is shown that
any probabilistic theory of the general type considered here, in which
every system arises as a non-signaling, locally tomographic product of
qubits, and in which, for every system $A$, the group $G(A)_{u}$ acts continuously on the set of pure states, is quantum. In contrast, our approach shows that, in the
context of a probabilistic theory in which systems are represented by Jordan models,  
the mere {\em existence} of a {\em single}
qubit, together with the possibility of forming locally tomographic,
non-signaling composite systems, is enough to enforce all the
structure of QM, including the aforementioned uniformity
assumption. We have a similar result for any dagger-HSD theory. 

Even so, various interesting questions remain regarding HSD
theories. For one thing, it would be very interesting to understand
the possibilities for {\em non}-locally tomographic composites in such
a theory: this should shed light on real and
quaternionic QM, in particular. In a different direction, one would want to know
whether it is possible to weaken, or entirely to dispense with, the assumption in
Theorem 1, that the theory $\cc$ includes a qubit. If so, then in the context of 
locally tomographic, non-signaling probabilistic theories in which systems are unital and uniform, {\em 
or} the context of dagger-monoidal probabilistic theories in which the canonical bilinear 
form is positive-definite, the HSD condition by itself is sufficient to rule out non-$C^{\ast}$-algebraic theories.   
Of course, it would be at least equally interesting to construct a non-$C^{\ast}$-algebraic, locally-tomographic, 
non-signaling HSD theory that {\em not} contain a qubit.

\noindent {\em Acknowledgement:} We thank C. M. Edwards for drawing
our attention to Hanche-Olsen's paper. Part of this work was done
while the authors were guests of the Oxford University Computing
Laboratory, whose hospitality is also gratefully acknowledged. H. B.
thanks the Foundational Questions Institute (FQXi) for travel support
for the visit.  Additional work was done at the Perimeter Institute
for Theoretical Physics; work at Perimeter Institute is supported in
part by the Government of Canada through Industry Canada and by the
Province of Ontario through the Ministry of Research and Innovation.\\

\begin{appendix}
{\bf Appendix: The Koecher-Vinberg Theorem}

{\small

This appendix contains a detailed sketch of the proof of the Koecher-Vinberg Theorem. This is almost entirely extracted from Faraut and Koranyi \cite{Faraut-Koranyi}, to whom we refer for many of the details,  but with a few minor modifications in order to obtain the precise form of the theorem (our Theorem 8) that we required above. 

In what follows, $\E$ is a finite-dimensional order-unit space with a
self-dual positive cone $\E_+$. By this we mean that {\em there
  exists} a self-dualizing inner product on $\E$. Let $\Aut(\E)$
denote the group of order-automorphisms of $\E$. This is a Lie group,
as is any closed subgroup $G \leq \Aut(\E)$.

When $G$ is connected and acts homogeneously on $\E_+$ (that is, transitively on the
interior of $\E_+$), we can use this action to construct a Jordan
product on $\E$, as per Theorem 8, which, for convenience, we now restate:

{\bf Theorem:} {\em Let $G$ be a closed, connected subgroup of
$\Aut(\E)$ and 
let $\g_u$ denote the Lie algebra of $G_u$, the stabilizer of $u$ in $G$.  Then
\begin{itemize} 
\item[(a)] It is possible to choose a self-dualizing inner product on $\E_+$ in such a way that 
$G_u = G \cap O(\E)$, where $O(\E)$ is the orthogonal group with respect to the chosen inner product;
\item[(b)] If $G = G^{\dagger}$ with respect to this inner product, then 
\[\g_u = \{ X \in \g | X^{\dagger} = -X\} = \{ X \in \g | Xu = 0\},\] 
and $\g = \p \oplus \g_u$, where $\p = \{ X \in \g | X^{\dagger} = X\}$. 
\item[(c)] In this case the mapping $\p \rightarrow \E$, given by $X
  \mapsto Xu$, is an isomorphism of linear spaces.  Letting $L_a$
  denote the unique element $X \in \p$ with $X u = a$, define
\[a \bullet b = L_{a} b\] 
for all $a, b \in \E$. Then $\bullet$ makes $\E$ a formally real Jordan algebra, with identity element $u$. 
\end{itemize} }

We break the proof into a series of Lemmas. Throughout, $G$ is a connected, closed subgroup of $\Aut(\E)$, acting 
homogeneously on the interior, $\E^{\circ}_{+}$, of the cone $\E_+$. 

{\bf Lemma A:} {\em If $g \in \Aut(\E)$, then $g^{\ast} \in \Aut(\E)$,
where $g^{\ast}$ is the adjoint with respect to any self-dualizing
inner product.}

{\em Proof:} If $g \in \Aut(\E)$ preserves $\E_+$, then $g^{\dagger}$ preserves $\E^{+} = \E_+$. $\Box$ 

{\bf Lemma B:} {\em Any compact subgroup of $\Aut(\E)$ fixes some point $a$
in the interior of $\E_+$. In particular, a maximal compact subgroup
{\em is} a stabilizer, and vice versa. Thus, all maximal compact
subgroups of $\Aut(\E)$ are conjugate.}

For a proof, see \cite{Faraut-Koranyi}, Proposition I.1.8ff.

{\bf Lemma C:} 
{\em For a suitable choice of self-dualizing inner product, $O(\E)
\cap G \leq G_u$, where $O(\E)$ is the orthogonal group relative to
the chosen inner product.}

{\em Proof:} If $\langle, \rangle$ is any inner product on $\E$ with
  respect to which $\E_{+} = \E^{+}$, one can show that $O(\E) \cap
  \Aut(\E)$ is the stabilizer of some $a \in \E_{+}^{\circ}$ 
  (\cite{Faraut-Koranyi}, Proposition I.1.9). It follows that $O(\E)
  \cap G \leq G_{a}$. Since $G$ acts transitively on $\E_{+}^{\circ}$,
  we can find some $g \in G$ with $ga = u$; replacing $\langle,
  \rangle$, if necessary, by the inner product $\langle x, y
  \rangle_{g} : = \langle g x, g y \rangle$ --- which is also
  self-dualizing, by \cite{Faraut-Koranyi}, Proposition I.1.7 --- we can
  assume that $a = u$, whence, that $O(\E) \cap G \leq G_u$. $\Box$

This gives us part (a) of the Theorem. 

Now let $K = G \cap O(\E)$, and let $\k$ denote $K$'s Lie algebra. Notice that $\k = \g \cap \mathfrak{o}(\E) = \{ X \in \g | X^{\dagger} = -X\}$. 

{\bf Corollary 1:} {\em Let the inner product on $\E$ be chosen as per Lemma C. Let $K = G \cap O(\E)$, and 
let $\k$ be the Lie algebra of $K$, and let $\g_u$ denote the Lie algebra of $G_u \leq G$. Then    
\begin{itemize} 
\item[(a)] $G_{u}$ is connected; 
\item[(b)] $G_{u} = K$; 
\item[(c)] $\g_u = \k$; 
\end{itemize} }

{\em Proof:} (a) $G/G_u$ is homeomorphic to the simply connected space $\E^{\circ}_{+}$; hence, as $G$ is connected, 
so is $G_{u}$ (see, e.g., \cite{Knapp}, Proposition 1.94). It follows that if $G_u$ and $K$ have the same Lieq algebra, they coincide --- in other words, (b) follows from (c).  To prove (c), note that since $K \le G_u$, by the choice of inner product, we have $\k \leq \g_u$. For every $X \in \g_u \leq \g$, we have 
the decomposition $X = X_1 + X_2$ with $X_1$ self-adjoint and $X_2$, skew-adjoint. Since $X_2 \in \k \subseteq \g_{u}$, it follows that $X_1 = X - X_2 \in \g_u$ as well, whence, $e^{tX_1} \in G_u$ for all $t$. However, $G_u$ is compact, so this implies that $e^{tX_1}$ is bounded as a function of $t$. Since $X_1$ is self-adjoint, this is possible only if $X_1 = 0$. Hence, $X = X_2 \in \k$, and we have $\g_u \leq \k$. $\Box$

{\em Proof of Part (b):} Suppose now that $G$ is self-adjoint with respect to the self-dualizing inner product chosen in Lemma C. Then, for every $X \in \g$, $X^{\ast} \in \g$. To see this, let $X = \gamma'(0)$ where 
$\gamma : {\mathbb R} \rightarrow G$ is a smooth path with $\gamma(0) = \1$, and note that $\gamma^{\ast} : t \mapsto 
\gamma(t)^{\ast}$ is another such path, with ${\gamma^{\ast}}'(0) = X^{\ast}$. 
It now follows that, for every $X \in \g$, $X_1 := (X + X^{\dagger})/2$ and $X_2 = (X - X^{\dagger})/2$ also lie in $\g$; we have $X = X_1 + X_2$, 
so that $\g$ decomposes as the direct sum $\g = \p + \k$, where $\p$ is the space of self-adjoint elements of $\g$ and $\k$, the space of skew-adjoint elements. This establishes part (b) of the Theorem. 

{\em Remark:} Note also, for later reference, that if $X, Y \in \p$, we also have $[X,Y] \in \g$ and $[X,Y]^{\dagger} = [Y,X] = -[X,Y]$, i.e., $[X,Y] \in \k$. 

The interior, $\E^{\circ}_{+}$,  of the cone $\E_+$ is a smooth manifold, on which $G$ acts smoothly. 
Thus, we have a canonical smooth mapping $\phi : G \rightarrow \E^{\circ}_{+}$ given by $g \mapsto gu$. Differentiating this, we obtain a linear mapping $d\phi (\1) : \g \rightarrow T_{u}(\E^{\circ}_{+}) = \E$. Explicitly, if $X = \gamma'(0) \in \g$, where 
$\gamma$ is a smooth path in $G$ with $\gamma(0) = \1$, then 
\begin{equation} d\phi(\1)(X) = \frac{d}{dt} \phi(\gamma(t))_{t = 0} = \gamma'(0)u = Xu.\end{equation}

{\bf Lemma D:} {\em Let $X \in \g$. Then $X \in \k$ iff $Xu = 0$.}

{\em Proof:} Suppose $d\phi(\1)(X) = Xu = 0$, where $X = \gamma'(0)$. The vector-valued function $v(t) = e^{tX}u$ then satisfies  
\[v' = Xv = Xe^{tX}u = e^{tX}Xu = 0.\] 
It follows that $v$ is constant, i.e, that $e^{tX}u = u$ for all
$t$. But then $e^{tX} \in G_u$, so that $X = \frac{d}{dt} e^{tX}|_{t =
  0} \in \k$. Conversely, if $X \in \k$, then $X = \gamma'(0)$ where
$\gamma(t) \in K$ and $\gamma(0) = \1$, so that $Xu = \gamma'(0) u =
[\frac{d}{dt}(\gamma(t)u)]_{t = 0} - [\gamma(t) \frac{d}{dt} u]_{t=0}
= [\frac{d}{dt}(\gamma(t)u)]_{t = 0} = \frac{d}{dt} u |_{t = 0} = 0$ (the
last equality uses $\gamma(t) \in K$ and $Ku=u$).  $\Box$

{\bf Corollary 2:} {\em $d\phi(\1) : \p \simeq \ran(d\phi(\1)) \leq \E$.} 

{\em Proof:} By Lemma D and (\theequation), $\k$ is the kernel of $d\phi(\1)$; as established above
(in the proof of part (b) of Lemma C, $\g = \p \oplus \k$. $\Box$ 

{\bf Lemma E:} {\em $\ran(d\phi(\1))$ is all of $\E$, i.e., $d\phi(\1) : \p \simeq \E$.}

{\em Proof:} Note, first, that $T_{g}(G) = gT_{1}(G) = g\g$ for any $g \in G$. We also have 
\[(d\phi(g))(gX) = gXu\]
for all $X \in \g$. Hence, $\ran(d\phi(g)) = g \ran(d\phi(1))$. If the latter is not all of $\E$, then 
every $g$ is a critical point of $\phi$, whence, every point $\phi(g) = gu \in \E$ is a critical value. 
Sard's Theorem now tells us that $Gu = \E^{\circ}_{+}$ has measure zero, a contradiction. $\Box$ 

{\em Construction of the Jordan product} Now define $L_x \in \p$ to be
the unique self-adjoint element of $\g$ with $L_x u = x$. Set $x
\bullet y = L_x y$ for all $x, y \in \E$. This is evidently
bilinear. A series of computations (see \cite{Faraut-Koranyi},
pp. 49-50) shows that it makes $\E$ a formally real Jordan algebra
with identity element $u$. Specifically,

(1) $x \bullet y = y \bullet x$ since $x \bullet y - y \bullet x 
= x \bullet (y \bullet u)u - y \bullet (x \bullet u) = [L_x,L_y]u = 0$ (as
remarked above following the proof of Corollary 1, $X, Y \in \p
\Rightarrow [X,Y] \in \k$, and, by Lemma D, $\k u = 0$);

(2) $x \bullet u = L_x u = x$ by definition of $L_x$, 
so $u$ serves as the identity; 

(3) The product satisfies the Jacobi identity. This is proved exactly
as in \cite{Faraut-Koranyi}. Note that the argument uses the fact that
the inner product is associative, which follows from $L_x$ being
self-adjoint. This also then gives us that the Jordan algebra $\E$ is
formally real.

This completes the proof of the Koecher-Vinberg Theorem. $\Box$ 

} \end{appendix} 

\end{document}